\documentclass[prf,aps,superscriptaddress,amsmath,amssymb,11pt,longbibliography]{revtex4-2}
\usepackage[latin1]{inputenc} 
\usepackage[T1]{fontenc}

\usepackage{graphicx}
\usepackage{framed}
\usepackage{esvect}
\usepackage{amsmath,amssymb,color,bm}
\usepackage{mathrsfs}
\usepackage[dvipsnames]{xcolor}
\usepackage{siunitx}

\newcommand\nocolor[1]{\textcolor{black}{#1}}
\newcommand\BC[1]{\textcolor{black}{#1}}
\newcommand\LB[1]{\textcolor{black}{#1}}

\newcommand\tb[1]{\textbf{#1}}

\newcommand\tn[1]{\textnormal{#1}}

\newcommand\mc[1]{\mathcal{#1}}

\newcommand\beq{\begin{equation}}
\newcommand\eeq{\end{equation}}
\newcommand\beqa{\begin{eqnarray}}
\newcommand\eeqa{\end{eqnarray}}

\newcommand\re[1]{\textnormal{Re}\left[#1\right]}

\newcommand\dd{\textnormal{d}}

\def\x{\tb{r}}
\def\q{\tb{q}}
\def\w{\omega}

\def\v{\tb{v}}

\def\F{\tb{F}}

\def\L{\mc{K}}

\def\.{\cdot}

\def\G{G_{\v}}
\def\F{\tb{F}_{\rm ext}}

\def\1{^{-1}}
\def\2{^{-2}}
\def\3{^{-3}}

\newcommand\pder[2]{\frac{\partial{#1}}{\partial{#2}}}

\begin{document}

\title{\nocolor{Modulation of} hydrodynamic permeability through fluctuating porous membranes}
\title{\nocolor{Hydrodynamic permeability of fluctuating porous membranes}}

\author{Albert Dombret}
\affiliation{Laboratoire de Physique de l'\'Ecole Normale Sup\'erieure, ENS, Universit\'e PSL, CNRS, Sorbonne Universit\'e, Universit\'e Paris Cit\'e, 24 rue Lhomond, 75005 Paris, France}
\author{Adrien Sutter}
\affiliation{Laboratoire de Physique de l'\'Ecole Normale Sup\'erieure, ENS, Universit\'e PSL, CNRS, Sorbonne Universit\'e, Universit\'e Paris Cit\'e, 24 rue Lhomond, 75005 Paris, France}
\affiliation{The Quantum Plumbing Lab (LNQ), \'Ecole Polytechnique F\'ed\'erale de Lausanne (EPFL), Station 6, CH-1015 Lausanne, Switzerland}
\author{Baptiste Coquinot}\email{Baptiste.Coquinot@ist.ac.at}
\affiliation{Laboratoire de Physique de l'\'Ecole Normale Sup\'erieure, ENS, Universit\'e PSL, CNRS, Sorbonne Universit\'e, Universit\'e Paris Cit\'e, 24 rue Lhomond, 75005 Paris, France}
\affiliation{Institute of Science and Technology Austria (ISTA), Am Campus 1, 3400 Klosterneuburg, Austria}
\author{Nikita Kavokine}
\affiliation{The Quantum Plumbing Lab (LNQ), \'Ecole Polytechnique F\'ed\'erale de Lausanne (EPFL), Station 6, CH-1015 Lausanne, Switzerland}
\author{Benoit Coasne}
\affiliation{ Univ. Grenoble Alpes, CNRS, LIPhy, F-38000 Grenoble, France}
\affiliation{ Institut Laue Langevin, F-38042
Grenoble, France}
\author{Lyd\'eric Bocquet}\email{lyderic.bocquet@ens.fr}
\affiliation{Laboratoire de Physique de l'\'Ecole Normale Sup\'erieure, ENS, Universit\'e PSL, CNRS, Sorbonne Universit\'e, Universit\'e Paris Cit\'e, 24 rue Lhomond, 75005 Paris, France}

\date{\today}

\begin{abstract}
In this paper we examine how {porosity fluctuations} affect \nocolor{the hydrodynamic} permeability \nocolor{of a porous matrix or membrane}. We introduce a fluctuating Darcy model, which couples the Navier-Stokes equation to the space- and time-dependent porosity fluctuations via a Darcy friction term. Using a perturbative approach, a Dyson equation for hydrodynamic fluctuations is derived and solved to express the permeability in terms of the matrix fluctuation spectrum. Surprisingly, the model reveals strong \LB{modifications} of the fluid permeability in fluctuating matrices compared to static ones. Applications to various matrix excitation models -- breathing matrix,  phonons,  active forcing -- highlight the significant influence of matrix fluctuations on fluid transport, offering insights for optimizing membrane design for separation applications.
\end{abstract}

\maketitle


\section{Introduction}

Porous media and membranes are the cornerstone of many industrial processes, from desalination to waste water treatment, from catalysis to energy storage and conversion. 
Fluid transport through these materials is generally a limiting factor, as the porosity hinders fluid motion. For example, a trade-off is unavoidable in filtration processes between the membrane selectivity  -- which is \nocolor{promoted by steric sieving} -- and its permeability -- \nocolor{limited by} the tortuous viscous flows across the porous matrix~\cite{Elimelech2017}. Bypassing such limitations is a key motivation to design new materials and principles for fluid transport~\cite{Werber2016,Gogotsi2019}.  Interestingly, nature has found strategies to partly circumvent such limitations, with the paradigmatic example of aquaporins -- both selective and highly permeable. Accordingly, the emerging properties of fluid transport at nanoscales, at the heart of nanofluidics, are definitely an asset in this quest~\cite{Siria2017,Faucher2019,Marbach2019}. A whole cabinet of curiosities has been unveiled, from nearly frictionless flows, dielectric anomalies, memory effects and non-linear ionic transport, to cite a few~\cite{bocquet2020,Robin2023a,aluru2023}.

Several studies suggested that fluctuations, whether in the fluid  or the confining material, are becoming an increasingly significant factor affecting nanoscale fluid transport~\cite{Bocquet2010,Davidovitch2005,Fetzer2007, Detcheverry2012}.
Wiggling channels were shown to play a significant role in biological transport ~\cite{Noskov2004, Bhabha2011, Wei2016}, as well as
in the  diffusion \nocolor{and separation} of species across artificial fluctuating channels~\cite{Zwanzig1992, Ma2015, Ma2016, Cruz-Chu2017, Marbach2018, kanduc2018, noh2021, Schlaich2024, ferreira2024}, while the flexibility of electrodes materials was shown to accelerate charging dynamics in supercapacitors~\cite{waysenson2025}. 
It is also noticeable that transport is not only affected by fluctuations in channel shape, but also by \nocolor{more complex collective modes of} the pore walls, \nocolor{such as} plasmons, phonons, etc~\cite{Kavokine2022, Bui2023, Coquinot2023b, Lizee2024}.
 \nocolor{Hence, this couples nanofluidics with the solid's degrees of freedom at the scale of a single pore~\cite{Coquinot2023,Coquinot2024,Coquinot2025}. }

In this paper, we investigate theoretically how fluctuations of a porous matrix influence \nocolor{its} hydrodynamic permeability. 
 The latter is defined as the averaged fluid velocity $\v_f$ under a pressure gradient  $\boldsymbol{\nabla} P$ 
\beq 
\langle \v_f\rangle =\frac{\L}{\eta}(-\boldsymbol{\nabla} P) 
\eeq 
with $\eta$ the fluid shear viscosity. The permeability has the dimension of a length squared. \nocolor{This is a collective transport property, which can  be interpreted, via fluctuation-dissipation theorem, as the collective diffusion of the fluid center of mass \cite{falk2015}.} 
\nocolor{For a static  structure, the bare permeability $\L_0$ accounts for the meandering flows  across the porous matrix. Now in a fluctuating matrix structure, 
not only the flow will be affected by the change in the typical pore size, but the breathing of the matrix porosity is expected to induce secondary flows that will impact the global dissipation, hence the permeability $\L$.} Molecular dynamics simulations have actually revealed that the impact of wall fluctuations is substantially larger on the collective diffusion as compared to any individual contribution~\cite{Marbach2018}. However, this counterintuitive effect remains so far unexplained.

\section{A fluctuating Darcy equation for the porous flow} 
\begin{figure*}[t]
    \includegraphics[width=\textwidth]{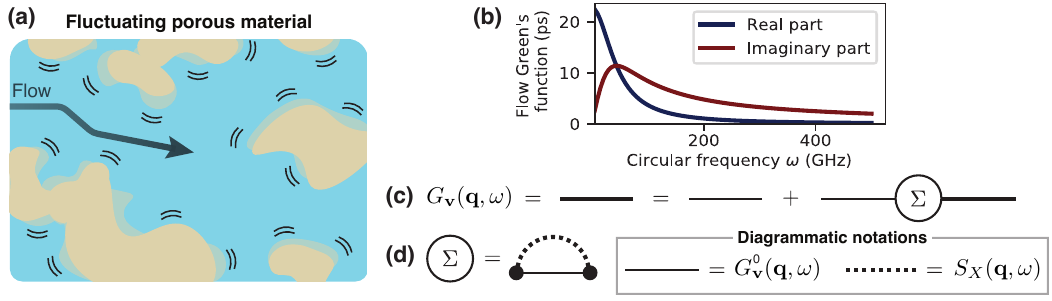}
    \caption{ \tb{Model and theoretical procedure.}
  \tb{(\nocolor{a})}   Schematic of the model: \nocolor{we consider a liquid flowing through a fluctuating porous medium}. \nocolor{At a coarse-grained level, the dissipation on the matrix is described in terms of a friction-like Darcy term in Navier-Stokes equation, here fluctuating in time and space.
      \tb{(\nocolor{b})} Green's function of the flow in absence of thermal fluctuations in the solid as a function of frequency for $q\approx6$ nm$\1$. 
    \tb{(\nocolor{c})} Diagrammatic Dyson equation to compute the effective Green's function $\G$ in presence of thermal fluctuations of the solid. These fluctuations are taken into account through a self energy $\Sigma$.
    \tb{(\nocolor{d})} Diagrammatic representation of the self energy $\Sigma$. 
 }
    }
    \label{fig1}
\end{figure*}

\nocolor{Fluid transport in porous membranes is usually well described by the coarse-grained Navier-Stokes-Darcy equation~\cite{Bear2018} and \nocolor{as a minimal model} we build on this framework to propose a fluctuating Darcy description to account for the matrix fluctuations}:
\begin{equation}\label{Darcy}
   \rho_m \pder{\mathbf{v}}{t} = - \boldsymbol{\nabla} P +  \eta \Delta \mathbf{v} -  {\rho_m}\xi(X) \mathbf{v} + \delta \mathbf{f}
\end{equation}
Here, $\mathbf{v}$ is the fluid velocity,  $\rho_m$ its mass density $P$ the pressure, and $\xi(X)$ the effective friction coefficient of the \nocolor{fluid} on the porous matrix according to Darcy's law; $X$ is a \nocolor{(fluctuating)} internal parameter characterizing the matrix porosity \nocolor{and its fluctuations}.
The friction coefficient is {\it de facto} defined at mesoscales, {\it i.e.} scales much larger than the pore sizes, and takes into account the friction encountered by the fluid while crossing the nanopores, see Fig.~\ref{fig1}a.
\nocolor{The non-linear advection term has been discarded since we are interested in low Reynolds-number flows taking into account fluctuations at frequencies and wavevectors where advection remains negligible.}
Finally, the fluctuations of the fluid dynamics originate from the Gaussian noise $\delta \mathbf{f}$, whose correlations are linked to the two dissipation terms -- the viscosity $\eta$ and the friction coefficient $\xi(X)$ -- according to the fluctuation-dissipation theorem.

The \nocolor{space- and time- dependent internal} parameter $X(\x,t)$ entering the friction coefficient $\xi(X)$ accounts for the internal microscopic degrees of freedom of the solid matrix and its porosity.
\nocolor{For example, below we will assume that $X$ identifies with the normalized fluctuations of the internal density of the breathing solid}. 
In the spirit of~\cite{Zwanzig1992}, we assume that this parameter fluctuates 
\nocolor{ slowly compared to the microscopic processes from which solid-liquid friction emerges -- typically above 100 THz for van der Waals and Pauli interactions -- but may reach frequencies comparable to hydrodynamic modes.}
\nocolor{We further assume that its fluctuations are independent from the hydrodynamic fluctuations, which is valid when the  modes of the solid either relax quickly to equilibrium or are dominated by an external forcing.}
\nocolor{Then in its most general form, it is a centred stochastic process characterised by its correlation functions, it particular its structure factor:}
\begin{equation}
     S_X(\x,t) =\langle X(\x,t) X(0,0)\rangle_0,
\end{equation}
\nocolor{where the subscript 0 refers to} \nocolor{averages taken on the solid matrix fluctuations} \nocolor{in the absence of interactions with the fluid.} 

\nocolor{The equations of motion are complemented by an equation of conservation of mass.
In our minimal model, we will restrict to fluctuations of the porous matrix geometry which modify the Darcy friction coefficient $\xi(X)$ but do no involve significant changes of volume -- like variations in pore sizes and solid's roughness.
As a consequence,  the equation of conservation of mass does not involve the variations of $X$ and can be standardly eliminated in Fourier space by applying the \BC{projector on transverse modes $\tb{J}(\q)=\tn{Id}-\q  \q/q^2$ on the induced velocity field.}
}

\nocolor{Finally, in order to illustrate the effects of the matrix fluctuations, we will consider a first-order expansion of the friction coefficient in $X$:}
\begin{equation}\label{exp-xi}
    \xi(X) = \xi_0+ \xi_1 X(\x,t)
\end{equation}
\nocolor{
Expressions for $\xi_1$ can be explicited for specific models and can be positive or negative.}
\nocolor{Beyond this approximation, one could consider higher order in the expansion of $\xi(X)$ in terms of the parameter $X$. Such corrections will affect the effect of fluctuations on the resulting permeability for strongly fluctuating matrices. 
We  leave such higher order terms to future work since we are mostly interested in the generic effects of fluctuations on the permeability. }

\section{Fluctuation-induced renormalization of the permeability} 

\subsection{Quasi-static regime} 
Let us first consider  the limiting regime where the variations of $X$ are much slower than the hydrodynamic fluctuations. 
\LB{The averaged permeability takes the expression  }
$ \L=\left\langle \frac{\nu}{\xi(X)}\right\rangle$ where the average is taken over $X$ and $\nu=\eta/\rho_m$ the kinematic viscosity. 
\LB{This result should be compared with the 'bare' permeability defined in terms of the quasi-static friction coefficient
and the average permeability 
$ \L_0=\frac{\nu}{\langle \xi(X)\rangle}$.
Assuming that $X$ follows a Gaussian law, and for the linear model $\xi(X) = \xi_0+ \xi_1 X(\x,t)$, one obtains 
${\L\over \L_0}\simeq 1+\left(\frac{\xi_1}{\xi_0}\right)^2 \langle X^2 \rangle $,which is valid in the limit of small fluctuations ({\it i.e.} neglecting terms beyond $\langle X^2 \rangle$).}
\LB{This quasi-static result is general by convexity and in the quasi-static limit, the average permeability is larger than the permeability obtained from the average friction:
$ \L\geq \L_0$.}
\LB{Note that this results remains valid for a quadratic friction model as well, with $\xi(X) = \xi_0+ \xi_1 X(\x,t)+\xi_2 X(\x,t)^2$.}

\subsection{General case} 
In the general case, the fluctuations of $X$ do couple with the hydrodynamic fluctuations described by the fluctuating Darcy equation, Eq. (\ref{Darcy}), and the quasi-static assumption no long\nocolor{er} holds. 
In order to calculate the momentum fluctuations and the permeability, one introduces
a generic deterministic external force $\rho_m\F(\x,t)$ applied to the fluid in Eq.(\ref{Darcy}) and computes the resulting velocity field. 
Going to Fourier space, we thus compute the Green's function $\G$ \BC{of transverse modes} of the \nocolor{velocity field} -- a generalized Stokeslet -- such that $\langle \v\rangle=\G\tb{J}\F$, \BC{with $\tb{J}$ the projector on transverse modes.}
The permeability $\L$ is then defined as the zero-frequency and zero-wavevector limit of the Green's function \BC{of transverse modes}:
\beq {\L\over\nu}=\G(\q=0,\omega=0) \label{K}\eeq
\nocolor{We note that the fluctuation-dissipation theorem allows rewriting Eq.(\ref{K}) in terms of a Green-Kubo relationship for $\L$, hence interpreted as a collective diffusion coefficient of the fluid center of mass fluctuations. 
}

\nocolor{We} now compute the Green's function of the flow in presence of the fluctuations of the solid. 
Using Eqs.~\eqref{Darcy}-\eqref{exp-xi} and incompressibility, the fluid velocity writes in Fourier space:
\begin{equation}\label{Darcyq}
 \mathbf{v}(\mathbf{q},\omega) =\G^0 \tb{J}\left[-\xi_1 X\!\circledast \!\v+ \delta\tb{f}+\F \right]
\end{equation}
where $\circledast$ denotes the convolution in both frequencies and wavevectors:
\beq [X\!\circledast \!\v](\mathbf{q},\omega)=\int\frac{\dd\bar\q\dd\bar\omega}{(2\pi)^4}X(\bar\q,\bar\omega)\v(\mathbf{q}-\bar\q,\omega-\bar\omega)\eeq
 Here, we have introduced the Green's function of the Darcy equation for a static solid: 
 \begin{equation}
  \G^0(\q,\omega)=  \frac{1}{q^2 \nu + \xi_0-i \omega}
\end{equation}
Depending on the wavevector, this non-interacting Green's function, \nocolor{shown in Fig.~\ref{fig1}b}, varies typically in the frequency range 10 GHz -- 100 THz.
If the solid's fluctuations are slow compared to the variations of hydrodynamic velocity, their convolution decouples
and we recover the quasistatic limit. 

However in the more general case, Eq.~\eqref{Darcyq} should be averaged over the various fluctuations to obtain the effective Green's function of the flow.
To this end, we introduce a perturbative expansion of the velocity field $\v$ in the coupling constants $\xi_1$.
\nocolor{At the leading order, we simply have
 \begin{equation}\label{v0}
 \mathbf{v}_0(\mathbf{q},\omega) =\G^0 \tb{J}\left[ \delta\tb{f}+\F \right]
\end{equation}
which we then inject in the right hand of Eq.~\eqref{Darcyq}, providing the basis of a systematic expansion in powers of $\xi_1$.
For instance, we obtain to second order:
 \begin{equation}
 \mathbf{v}_2(\mathbf{q},\omega) =\xi_1^2\G^0 \tb{J} X\circledast\left(\G^0\tb{J}  X
\!\circledast \!\left(\G^0 \tb{J}\left[ \delta\tb{f}+\F \right]\right)\right)
\end{equation}
We then average over thermal noise using that $\langle X\rangle=0$, $\langle \delta\tb{f}\rangle=0$ and  $\langle X \delta\tb{f}\rangle=0$. 
We assume $X$ to be a Gaussian field with correlations 
\beq \langle X(\mathbf{q},\omega)X(\bar\q,\bar\omega)\rangle=(2\pi)^4\delta(\omega+\bar\omega)\delta(\q+\bar\q)S_X(\mathbf{q},\omega) \eeq  
As a consequence, the odd orders of the expansion vanish in average and the first relevant term is:
 \begin{equation}\label{v2}
 \langle\mathbf{v}_2(\mathbf{q},\omega)\rangle =\G^0\tb{J}\boldsymbol{\Sigma}(\mathbf{q},\omega)\G^0 \tb{J}\F
\end{equation}
where we have introduced the self-energy  
\beqa
\Sigma(\q,\omega) &=& \xi_1^2\, \left[S_X\!\circledast\!\alpha_\tb{J} G^0\right](\q,\omega)
\label{Self}
\eeqa
Note that we introduced a factor $\alpha_\tb{J}$, which is a geometrical factor originating in the projections \BC{on transverse modes}. As shown in  Appendix A, this leads {\it in fine} to a modification of the prefactor, calculated as 5/6. } 

\nocolor{Beyond the second order, this expansion can be made systematic thanks to a Dyson equation: }
\beq \G(\q,\omega)=\G^0(\q,\omega)+\G^0(\q,\omega)\Sigma(\q,\omega)\G(\q,\omega)\eeq
\nocolor{which is represented diagrammatically in Fig.~\ref{fig1}c-d.}
After resummation, the average fluid velocity then reads $ \langle\mathbf{v}\rangle =\G \tb{J}\F$ with an effective Green's function:
\begin{equation}\label{Green}
 \G(\q,\omega)=\frac{1}{q^2 \nu + \xi_0-\Sigma(\q,\omega)-i \omega}
\end{equation}
Physically, it corresponds to the Fourier transform of the effective Darcy equation describing the flow after averaging \nocolor{over the fluctuations of the solid}.

We can now expand the self energy $\Sigma$ in powers of the wave-vector $\q$ (there is no linear term in $\q$ due to isotropy), writing
\beq \Sigma(\q,\omega) = \Sigma(q=0,\w)+ \frac{1}{2}\partial_q^2\Sigma(q=0,\omega) q^2 +\ldots \eeq
which allows calculating effective \nocolor{parameters} for the Darcy equation in the presence of the \nocolor{matrix fluctuations}. 
At zeroth order, we can interpret the self-energy as a correction to  Darcy's friction coefficient.
{\nocolor{In the present modelling,} the apparent friction is always decreased} 
as compared to the static \nocolor{case} $\xi_0$: $\xi_{\rm app}=\xi_0-\Delta \xi$, with $\Delta \xi = \Sigma(\q=\tb{0},\omega=0) >0$ calculated as 
\beq \label{xi}
\Delta \xi = \frac{5\xi_1^2}{12\pi^3}\,\int_0^\infty\dd q\int_{\nocolor{0}}^{\nocolor{\infty}}\dd\omega\,q^2 S_X(q,\omega)\re{\G^0(q,\omega)}
\eeq
The renormalized permeability of the system is then deduced as 
 \beq \label{GK2}
  \L=\frac{\nu}{\xi_{\rm app}}= {\L_0\over1-\frac{\Delta\xi}{\xi_0}}
  \eeq
  with $\L_0=\nu/\xi_0$ the static permeability. 
\nocolor{Quantitatively, we find that in all practical cases (see below), the permeability cannot be described by the quasi-static result, confirming the need for tackling fluctuations in full generality.}
We finally note that, similarly to the quasistatic case, \nocolor{non-Gaussian terms have to be accounted for large fluctuations (hence large $\Delta \xi$),  }
which can be handled  \nocolor{systematically}  as additional higher-order self-energy \nocolor{terms}. \nocolor{We leave this detailed calculation 
for future studies and focus here on the qualitative effect of fluctuations on the permeability.}

\nocolor{Equations (\ref{xi})-(\ref{GK2}) are the main result of this work. They predict the renormalized permeability in terms of the solid fluctuation spectrum, $S_X$.}
Strikingly, \nocolor{under the present assumptions of the model,} $\Delta\xi$ is systematically positive and thus the permeability is always enhanced \nocolor{in the presence of fluctuations}. 
\nocolor{Furthermore, according to Eq.~\eqref{xi}, the increase in permeability will be maximized when the spectra of the hydrodynamic and solid-state fluctuations do overlap, corresponding to a syntonic frequency matching.}

\nocolor{Going now to second order in $\q$ in the expansion of $\Sigma$, we find a term that}
can be interpreted as a correction to the viscosity originating from the \nocolor{coupling of solid correlations} to the fluid dynamics. We obtain accordingly an apparent viscosity
$\nu_{\rm app}=\nocolor{\nu}+\Delta \nu$, with \beq \label{nu}
\Delta \nu =\frac{5 \xi_1^2}{6}\,\int_0^\infty\frac{\dd q\dd\omega}{4\pi^3}\,q^2 S_X(q,\omega)\partial_q^2\re{\G^0(q,\omega)}
\eeq
However, in spite of its fundamental interest, we find that this viscosity correction is negligible is most practical situations, see Appendix A.

Altogether, gathering the contributions of the \nocolor{apparent} permeability and viscosity, one {obtains} a \nocolor{renormalized} Darcy equation 
which \nocolor{accounts for fluctuations of the porous matrix}. 
Let us now apply this formalism to practical situations in order to evaluate the effects on the permeability. This requires 
\nocolor{specifying} the structure factor $S_X$ of the \nocolor{matrix} fluctuations.


\begin{figure*}
    \centering
    \includegraphics[width=\textwidth]{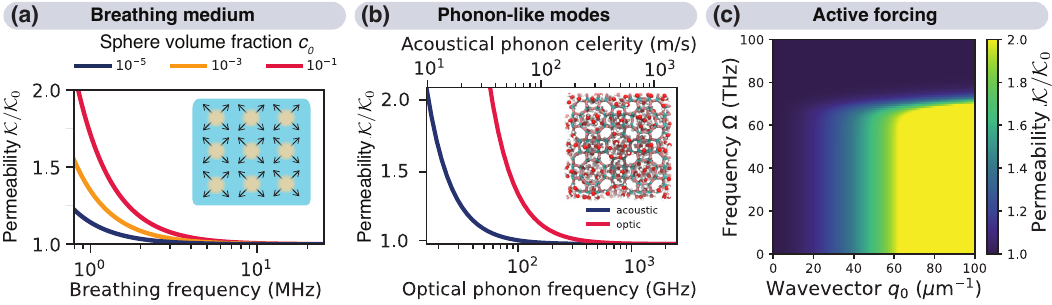}
    \caption{ \tb{Fluctuation-induced permeability modification $\L/\L_0$ for various fluctuation spectra $S_X$ of the matrix.}
    \tb{(a)} Across a breathing medium with sphere volume fraction $c_0$  as a function of the characteristic frequency $\omega_0/2\pi$. Inset: schematic of the model.
     \tb{(b)} Across a fluctuating porosity with phonon-like modes as a function of the characteristic frequency $\omega_0/2\pi$ for dispersionless optical phonons (bottom axis) and as a function of the {sound velocity} $c$ for propagating acoustic phonons (top axis). Inset: Molecular Dynamics image of water in a zeolite.
     \nocolor{In \tb{(a)} and \tb{(b)}, the solid fluctuations are thermal with temperature $T= 300$ K.}
     \tb{(c)} Across an actively forced solid matrix as function of the forcing frequencies $\Omega$ and wavevector $q_0$.
    }
    \label{fig2}
\end{figure*}



\section{Application to fluctuating matrices}

\subsection{Permeability across a breathing array of spheres}

As a first, prototypical example, we  consider a porous matrix made of an array of  fluctuating spheres, with  radius $R$ and a number density $\rho_s$. The liquid can flow through the array, with a Reynolds number assumed to be much lower than 1. 
In the static case, the permeability $\L$ can be explicitly calculated as a function of the sphere radius $R$ and the volume fraction of the spheres $c = \frac{4}{3}\pi R^3 \rho_s$ \cite{sangani1982}. 
Indeed, the force exerted on the flow by one sphere of the array can be computed as 
\begin{equation}
	F_{\rm 1} = 6 \pi \eta R \kappa(c)v,
\end{equation}
where the function $\kappa(c)$ takes the following form for $c \ll 1$:
\begin{equation}
	\kappa(c) \simeq 1 + \alpha c^{1/3}, \quad \textrm{with} \; \alpha \simeq 1.7601.
\end{equation}
Then, the volumic friction force that the fluid undergoes when flowing through the medium reads:
\begin{equation} \label{frictionforce}
	f = \rho_s F_1 = 6\pi \eta \rho_s R \kappa(c)v.
\end{equation}

\nocolor{
Let us now assume that each sphere has some intrinsic \emph{breathing} dynamics, leading to fluctuations of the radius $R = R_0 + r$ of the sphere around a mean value $R_0$, \BC{and that these dynamics are uncorrelated between spheres.}
One may accordingly define the parameter $X$ for a sphere as the normalized variation of the inner density of the spheres, {\it i.e.} $X\equiv (R_0/R)^3-1\simeq -3 r/R_0$. The volume fraction $c$ then behaves as $c=c_0(1+3 r/R_0)=c_0 (1-X)$
Thus, at lowest order in $X$, 
\begin{equation}
	f = \rho_m(\xi_0 + \xi_1 X) v,
\end{equation}
with \begin{equation} \label{xibreath}
	\xi_0 = \frac{6 \pi \eta a\rho_s}{\rho_m} \left(1 + \alpha c_0^{1/3} \right) \quad \textrm{and}  \quad \xi_1 =  -\frac{2 \pi \eta a\rho_s}{\rho_m} \left(1 +  2 \alpha c_0^{1/3} \right).
\end{equation}
}

\nocolor{
We now specify the breathing dynamics for $r$, which we model by an overdamped Langevin equation on $r$.
We denote by $\omega_0$ the intrinsic breathing frequency of the sphere, and $\gamma$ the damping rate of the \emph{breathing} modes, such that the fluctuating dynamics of the sphere reads 
\beq
\gamma \dot{r} = - \omega_0^2 \, r + \frac{1}{m}\delta f(t), 
\eeq
with $m$ an apparent mass of the radius dynamics and $\delta f(t)$ being some Gaussian noise with zero average. We can then directly compute the fluctuations of the sphere radius, embedded in its structure factor $S_r(t-t') = \langle r(t)r(t') \rangle$. The Fourier transform of $S_r$ is found by either solving the Langevin equation or using the fluctuation-dissipation theorem:
\begin{equation}
	S_r(\omega) = \frac{2 \gamma k_B T}{m} \frac{1}{(\gamma \omega)^2 + \omega_0^4}.
\end{equation}
}

 The fluctuations of the parameter $X$ are accordingly characterized by a structure factor $S_X(\mathbf{q}, \omega) = \frac{9}{R_0^2} S_r(\mathbf{q}, \omega)$. 
 \BC{Since the sphere dynamics are uncorrelated, the structure factor is uniform in wavevectors until reaching $q_{\rm max}\sim \rho_s^{1/3}$, associated with the typical inter-sphere distance. }
 Hence, we obtain the following structure factor of the internal density of the breathing spheres $X$:
\BC{\begin{equation}
	S_X^{\rm breath} (\mathbf{q}, \omega) \approx \frac{18 \gamma k_B T}{m R_0^2} \frac{1}{(\gamma \omega)^2 + \omega_0^4} \frac{(2\pi)^3}{V_q} \theta ( q_{\rm max}-q).
\end{equation}
where $\theta$ is the Heaviside distribution and  $V_q=\frac{4\pi}{3}q_{\rm max}^3$. }

We can then compute analytically the correction to the friction coefficient from Eqs.~(\ref{xi})-(\ref{GK2})  as:
\BC{\beq
	\Delta \xi 
	= \frac{45\rho_s \xi_1^2 k_B T}{2 m \w_0^2 \nu q_{\max}^2 R_0^2 }\left[ 1-\frac{\arctan\!\left(q_{\max}\ell(\w_0)\right)}{q_{\max}\ell(\w_0)}\,\right],
\eeq
where 
\beq  \ell(\w_0)^2 = \frac{\nu}{\xi_0+\omega_0^2/\gamma} \eeq
is a viscous diffusive length. 
Finally, the renormalized permeability of the system simply rewrites in the case of small fluctuations $\langle r^2 \rangle=k_BT/m\w_0^2$:
\begin{equation}
	\frac{\mathcal{K}}{\mathcal{K}_0} \simeq 1 + \frac{45 \xi_1^2}{2 \xi_0 \nu q_{\max}^2 }\, \frac{\langle r^2 \rangle}{R_0^2}\left[ 1-\frac{\arctan\!\left(q_{\max}\ell(\w_0)\right)}{q_{\max}\ell(\w_0)}\,\right]\label{sphere_breathing_eq}
\end{equation}}
Results for the breathing frequency dependence of the permeability are shown in Fig.~\ref{fig2}a.
For $\omega_0\rightarrow 0$, the excess permeability is directly proportional to the RMS of the matrix radius as $\Delta \L / \L_0\propto \langle r^2 \rangle / R_0^2$,  recovering the quasi-static result as expected. 
\BC{For large $\omega_0$ such that $q_{\max}\ell(\w_0)\ll 1$, Eq. \eqref{sphere_breathing_eq} reduces to:
\begin{equation}
	\frac{\mathcal{K}}{\mathcal{K}_0} \simeq  1 + \frac{15\xi_1^2}{2\xi_0^2} \frac{\langle r^2 \rangle}{R_0^2}\frac{1}{1+\w_0^2/\gamma \xi_0}, \label{sphere_breathing_eq0}
\end{equation}
which decays like $1/\w_0^2$. For  $\w_0\rightarrow \infty$,  the dynamics of the liquid no longer couples with the hydrodynamics and the permeability enhancement vanishes. }

\subsection{Permeability across a fluctuating porosity with phonon-like modes}

\nocolor{Beyond the previous simple model, the solid matrix dynamics usually exhibit more complex dynamical features, such as (optical or acoustic) phonons.}
\nocolor{Acoustic phonons describe propagating waves, while optical phonons are 
associated with (nearly) dispersionless breathing.} 
Phonon modes are rooted in the dynamics of the displacement field of the porous matrix, say $\mathbf{u}$, and the parameter $X$ is accordingly defined as the normalized density of the matrix as $X=-\nabla\cdot \mathbf{u}$. It is now a time- and space-dependent fluctuating quantity. 
\nocolor{
The fluctuation of the density will affect accordingly the pore size of the porous medium, hence the Darcy friction. As a rule of thumb, 
let us consider a dense porous material with a typical pore size $R$. The fluctuation of the density will therefore affect the pore radius as $R=R_0(1+{1\over 3}X)$. The Darcy friction stems from the viscous flow inside the porosity 
and should accordingly scale as $\xi \propto 1/R^2$.  The friction thus writes $\xi(X)=\xi_0+\xi_1 X$ with $\xi_1=-(2/3)\xi_0$ in this case.
In a porous material with a more complex structure, coefficients may differ but a similar linear expansion of $\xi(X)$ is expected. }

\nocolor{We now turn to the structure factor and first consider acoustic phonons with sound velocity $c$.
We consider a classical phonon description in terms of springs separating atoms of mass $m$ with drag coefficient $\gamma$, and the dynamical equation governing the displacement field $\mathbf{u}$ reads in the continuous limit
\beq
m \partial_t^2 \mathbf{u} = - m \gamma \partial_t \mathbf{u} + mc^2 \nabla^2 \mathbf{u} - \nabla V_{\rm ext}.
\eeq
with $V_{\rm ext}$ an external potential. 
Hence going to Fourier space, we deduce the response function for $X$ as
\beq
X(\mathbf{q}, \omega) = \frac{1}{m} \frac{q^2}{\omega^2 - q^2 c^2 + i \gamma \omega} V_{\rm ext}(\mathbf{q}, \omega).
\eeq
and the fluctuation-dissipation theorem gives
\beq
S_X^{\rm ph}(\mathbf{q}, \omega) = \frac{2 k_B T m}{\omega\rho^s_m } \text{Im} \left( \frac{1}{m} \frac{q^2}{\omega^2 - \omega_q^2 + i \gamma \omega} \right).
\eeq
with $\omega_q = c\,q$ for acoustic phonons and $\rho_m^s$ is the global mass density of the solid. 
Taking the limit $\gamma \rightarrow 0$ leads us to the simplified form for the structure factor associated with phonon-like excitations:
\beq
S_X^{\rm ph}(\mathbf{q}, \omega) = \frac{\pi k_B T q^2}{\rho^s_m \omega^2}  \delta (\omega \pm \omega_q) .
\eeq
In the following we will assume that the wavevector $q$ is limited by the interatomic distance $a$ i.e. $q \leq q_{\rm max} = 2 \pi / a$. 
For optical phonons, one finds the same expression except that the dispersion relation is now $\omega_q = \omega_0$, where $\omega_0$ is the constant optical phonon frequency.}

\nocolor{In the quasistatic limit, one can deduce 
the permeability deviation $\Delta \L / \L$ in terms of the RMS of the parameter $X$. The latter is given by 
\beq 
\langle X^2 \rangle = \int_0^\infty \dd \omega \int_0^{q_{\rm max}} \dd q \, q^2 \, S_X^{\rm ph}(q, \omega).
\eeq
A straightforward analytical evaluation of these integrals leads to the quasi-static estimate for acoustic phonons:
\beq \label{quasistat_ph}
\L\simeq \L_0\left(1+\frac{2 q_{\rm max}^3}{27\pi^2\rho_m^s c^2}k_B T\right)
\eeq
}
Noteworthy, we find a permeability enhancement which scales with $1/c^2$. 
\nocolor{This is originates in the contribution of lower energy phonon states which are more populated at a given temperature.}

Now beyond the quasistatic regime, {Eqs.~\eqref{xi}-\eqref{GK2} allows us to calculate the renormalized permeability  $\L/\L_0$ using the full expression for the $X$ spectrum. Results for the permeability as a function of the phonon frequency are shown in Fig.~\ref{fig2}b :
{the permeability is plotted as a function of the frequency $\omega_0$ for the optical phonons (bottom axis) and
as a function of the sound velocity $c$ for the acoustic phonons (top axis)}. 
In both cases, we find a strong permeability enhancement for soft solid matrices, associated with low frequency modes or low sound velocity.
\nocolor{While this follows the same trend as in the quasi-static limit, we find important quantitative corrections in the general case due to the dynamical contribution of the overlap between the phonon modes of the matrix and the hydrodynamic modes which do strongly increase the permeability. 
}

\subsection{Permeability across an actively forced solid matrix} 
Finally, \nocolor{the solid may be subject to} an active external forcing at a frequency $\Omega$ and (isotropic) spatial wavevector $q_0$. 
The effective structure factor of the fluctuations  writes accordingly
\begin{equation}
    S_X^{\rm act} ( \mathbf{q}, \omega) = (2 \pi )^2 \frac{A}{\rho_s^{2/3}} \delta (\omega - \Omega) \delta(q-q_0)
\end{equation}
\nocolor{where $\rho_s$ is the solid particle density and $A$ is an dimensionless amplitude of the forcing.} 
The resulting permeability enhancement is shown in Fig.~\ref{fig2}c for \nocolor{a range of} frequencies and wavevectors of the external active forcing (the value $\xi_1=-(2/3)\xi_0$ was assumed for simplicity).
\nocolor{We observe a permeability enhancement in a broad region of the frequency-wavevector space}. 
\nocolor{The amplification is maximum for large $q_0$, for which }
the viscous frequency $\nu q_0^2$ \nocolor{dominates over} the bare friction $\xi_0$. 
\nocolor{In contrast, the effect disappears for large frequency $\Omega$, as }
 the fluid \nocolor{decouples from} the forcing. 
\nocolor{Altogether, it appears possible to enhance the permeability of a porous matrix through an external forcing with well-chosen wavelength and frequency}. 

\section{Conclusion}
Using perturbation theory of a fluctuating Darcy equation, we have shown that 
the \nocolor{matrix} fluctuations renormalize the fluid permeability \nocolor{as they couple to} hydrodynamic modes.  In a counterintuitive way, 
this renormalization effect \nocolor{increases} the flow permeability in the models under scrutiny. 
\nocolor{Our model is minimal in terms of assumption and merely serves as an illustration to show the effect of matrix fluctuations. }
\nocolor{While corrections to our minimal model may reduce the effect -- and perhaps even lead to a permeability reduction in some cases  -- our results nevertheless reveal that fluctuations of a porous matrix can modulate significantly its permeability to fluid flow and provides insights in the phenomenology and ingredients at play.}
Indeed, we found that the permeability enhancement is optimal  
\nocolor{in the case of frequency matching}
between the modes of the solid and the hydrodynamic fluctuations of the fluid. 
\nocolor{By exploring elementary models of solid that describe their thermal fluctuations (breathing, phononic), we also provided hints as to the link between the microscopic characteristics of the porous matrix and its hydrodynamic permeability}. 
Furthermore, we have shown that an external forcing could also be used to control and increase the permeability by exciting \nocolor{well-chosen} modes in the solid.
\nocolor{For specific systems, numerical Molecular Dynamics simulations that account for solid fluctuations have successfully explored the modulation of ionic diffusion, and could be extended to study mass transport, providing valuable quantitative estimates of permeability modulation. }

\nocolor{Beyond our model, it would be interesting to extend such formalism to account for a more general framework of the solid's fluctuations, including coupling to the liquid's volume, correlation with hydrodynamic fluctuations and non-Gaussian correlations. 
In terms of systems,} it would be also interesting to investigate the individual and collective motion of penetrants in fluctuating \nocolor{anisotropic} (polymer or glass-forming) matrices
\cite{zhang2017,kanduc2018}.

\nocolor{Overall, our results provide a new direction of research for nanofluidics and suggest new strategies for improving the efficiency of membranes.  
Indeed, exploiting membrane fluctuations seems to be a promising and} unexplored lever to circumvent the permeability-selectivity trade-off that hinders membrane-based separation technologies.

\newpage

\section*{Appendix A: Perturbation theory for the renormalised permeability and viscosity}

We provide some details of the calculation described in section III.B. 

As shown in Section III.B, the self-energy takes the expression
\beqa\label{v2a2}
\boldsymbol{\Sigma}(\q,\omega) &=& \xi_1^2\,\int\frac{\dd\bar\q\dd\bar\omega}{(2\pi)^4}\, \G^0 (\bar\q,\bar\omega) S_X(\q-\bar\q,\omega-\bar\omega) \tb{J}(\q)\tb{J}(\bar\q)\tb{J}(\q)\eeqa
where we recall that the projector matrix is defined as $\tb{J}={\rm Id} -\q\q/q^2$.
Note that we have used the identity $\tb{J}^2=\tb{J}$. The self energy is in general a matrix in spatial coordinates.

We can simplify this expression by noting that the self energy reduces to a multiple of $\tb{J}(\q)$ for isotropic systems. 
Indeed, we can define spherical coordinates around the direction of $\q$ and explicit the effect of the self energy matrix on a generic vector $\tb{w}=(w\quad \theta_w\quad 0)_{\rm sph}$.
Here we have fixed $\varphi$ according to the direction of $\tb{w}$. 
We denote $\bar \q = (\bar q \quad \bar\theta\quad \bar\varphi)_{\rm sph}$.
Then, in the associated Cartesian coordinates we have 
\beq  \tb{J}(\q)\tb{J}(\bar\q)\tb{J}(\q)\tb{w} = w\cos(\theta_w)\times (0\quad 1-\cos(\bar\theta)^2 \cos(\bar\varphi)^2\quad -\cos(\bar\theta)^2\cos(\bar\varphi)\sin(\bar\varphi) )_{\rm Cart} \eeq
Since the system is isotropic, $\G^0$ and $S_X$ only depend on the norms of $\bar \q$ and $\q-\bar \q$. 
In particular, they do not dependent on $\bar \varphi$. 
Thus averaging on $\bar\varphi$ we obtain:
\beq \langle \tb{J}(\q)\tb{J}(\bar\q)\tb{J}(\q)\tb{w}\rangle_{\bar\varphi} = w\cos(\theta_w)\times (0\quad 1-\frac{1}{2}\cos(\bar\theta)^2\quad0) )_{\rm Cart} =\left(1-\frac{1}{2}\cos(\bar\theta)^2 \right) \tb{J}(\q)\tb{w}  \eeq
Therefore, the projector $ \tb{J}(\q)$ being already present in Eq.~\eqref{v2}, we can see the self energy as a scalar function:
\beq
\Sigma(\q,\omega) = \xi_1^2\,\int\frac{\dd\bar\q\dd\bar\omega}{(2\pi)^4}\, \left(1-\frac{1}{2}\left(\frac{\q\cdot\bar\q}{q\bar q}\right)^2 \right)\G^0 (\bar\q,\bar\omega) S_X(\q-\bar\q,\omega-\bar\omega)\eeq
Relaxing the incompressibility assumption, we would have a similar result with a more complicated and frequency-dependent additional geometrical factor.
Now, for $q\ll \bar q$ the isotropic function $S_X(\q-\bar\q,\omega-\bar\omega)$ no longer depends on the angle $\q\cdot\bar\q$ and the geometrical factor $1-\frac{1}{2}\left(\frac{\q\cdot\bar\q}{q\bar q}\right)^2$ reduces to $\frac{5}{6}$ upon integration.

As shown in the main text, 
the renormalised Green's function of the flow then takes the expression
\beq \G(\q,\omega)=\frac{1}{\G^0(\q,\omega)^{-1}-\Sigma(\q,\omega)}=\frac{1}{q^2 \nocolor{\nu} + \xi_0-\Sigma(\q,\omega)-i \omega}\eeq

By isotropy, we can expand the self energy as a function of $q$:
\beq \Sigma(\q,\omega) = \Sigma(q=0,\w)+ \frac{1}{2}\partial_q^2\Sigma(q=0,\omega) q^2 +\ldots \eeq
The quadratic term provides the correction to the viscosity at small frequency.
Note that the geometrical factor $1-\frac{1}{2}\left(\frac{\q\cdot\bar\q}{q\bar q}\right)^2$ depends only on the angle between $\q$ and $\bar\q$ but not on the norm $q$. 
Thus, we obtain:
\beq\label{v2av}
\Delta\xi=\Sigma(0,0) = \frac{5}{6}\xi_1^2\,\int_0^\infty\frac{\dd q\dd\omega}{2\pi^3}\,q^2 S_X(q,\omega)\re{\G^0(q,\omega)}
\eeq
\beq
\Delta\eta=\frac{1}{2}\partial_q^2\Sigma(q=0,0) = \frac{5}{6}\xi_1^2\,\int_0^\infty\frac{\dd q\dd\omega}{4\pi^3}\,q^2 S_X(q,\omega)\partial_q^2\re{\G^0(q,\omega)}
\eeq
The resulting viscosity deviation vanishes for the model of breathing, which happens at $\q=0$,  and is small for the other models as shown in Fig. \ref{fig3}.

\begin{figure*}
    \centering
    \includegraphics[width=\textwidth]{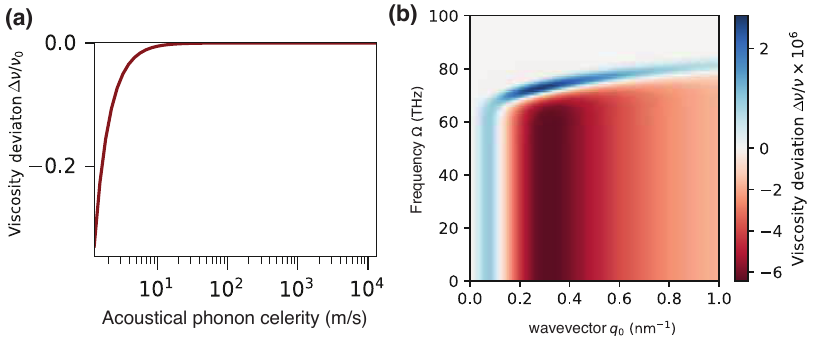}
    \caption{\tb{Fluctuation-induced viscosity deviation $\Delta\nu/\nu_0$ for various fluctuation spectra $S_X$ of the matrix.}
       \tb{(a)} Across a fluctuating porosity with phonon-like modes as a function of the sound velocity $c$ for propagating acoustic phonons. The solid fluctuations are thermal with temperature $T= 300$ K.
     \tb{(b)} Across an actively forced solid matrix as function of the forcing frequencies $\Omega$ and wavevector $q_0$.
    }
    \label{fig3}
\end{figure*}

\newpage

\section*{Appendix B: Numerical computations}

The numerical parameters used to obtain the plots given in the main text are summarized in table ~\ref{tableau}. 

\begin{table}[h]
\centering
\begin{tabular}{|c|c|}
\hline
Liquid &  $T = 300\;  \si{K}$ \\
& $\nu = 10^{-6}\;  \si{m^2.s^{-1}}$ \\
& $\rho_m = 10^{3}\;  \si{kg/m^{3}}$ \\
\hline
Breathing & $R_0 = 1\; \si{ nm}$ \\
spheres & $c = \frac{4}{3}\pi R^3 \rho_s$ \\
& $q_{\rm max} = \rho_s^{1/3}$ \\
& $\gamma = 10^6\;  \si{s^{-1}}$ \\
& $m = 10^{-14}\;  \si{kg}$ \\
& $ \alpha = 1.7601$ \\
& $\eta = 10^{-3}\;  \si{Pa.s}$\\
\hline
Phonons & $\xi_0 = 4 \cdot 10^{10} \si{s^{-1}}$ \\
& $\xi_1 = - \frac{2}{3} \xi_0$ \\
& $\rho_m^s = 2.25 \cdot 10^3 \si{kg . m^{-3}}$ \\
& $ q_{\rm max} = 2 \pi / a$, $a = 1$\,\AA \\

\hline
Forcing & $\rho_s = 10^{18}\;  \si{m^{-3}}$ \\
& $A = 10^{-3}$ \\
\hline
\end{tabular}
\caption{Parameters used in numerical computations.}\label{tableau}
\end{table}

\newpage


\section*{Acknowledgements} 
The authors acknowledge support from ERC project {\it n-AQUA}, grant agreement $101071937$. B.C. and A.S. acknowledge support from the CFM Foundation. B.C. acknowledges support from the NOMIS Foundation.

\bibliography{bibfile}

\end{document}